\documentclass{article}

\usepackage[preprint]{neurips_2025}

\usepackage{multirow}
\usepackage[utf8]{inputenc} 
\usepackage[T1]{fontenc}    
\usepackage{hyperref}       
\usepackage{url}            
\usepackage{booktabs}       
\usepackage{amsfonts}       
\usepackage{nicefrac}       
\usepackage{microtype}      
\usepackage{xcolor}         
\usepackage[many]{tcolorbox}
\usepackage{lipsum}
\usepackage{graphicx}
\usepackage{amsmath}
\usepackage{cleveref}
\usepackage{caption}
\usepackage{float}      

\usepackage[linesnumbered, ruled, vlined]{algorithm2e}

\title{A Mixture of Linear Corrections Generates\\ Secure Code}

\author{
  Weichen Yu$^{1}$, Ravi Mangal$^{2}$, Terry Zhuo$^{3}$,  Matt Fredrikson$^{1}$, Corina S. Pasareanu$^{1}$\\
  \\$^1$Carnegie Mellon University \quad $^2$Colorado State University $^3$Monash University \\ 
  \texttt{wyu3@andrew.cmu.edu} \\
}

\begin{document}

\maketitle

\begin{abstract}

Large language models (LLMs) have become proficient at sophisticated code-generation tasks, yet remain ineffective at reliably detecting or avoiding code vulnerabilities. Does this deficiency stem from insufficient learning about code vulnerabilities, or is it merely a result of ineffective prompting? Using representation engineering techniques, we investigate whether LLMs internally encode the concepts necessary to identify code vulnerabilities. We find that current LLMs encode precise internal representations that distinguish vulnerable from secure code--achieving greater accuracy than standard prompting approaches. Leveraging these vulnerability-sensitive representations, we develop an inference-time steering technique that subtly modulates the model's token-generation probabilities through a mixture of corrections (MoC). Our method effectively guides LLMs to produce less vulnerable code without compromising functionality, demonstrating a practical approach to controlled vulnerability management in generated code. Notably, MoC enhances the security ratio of Qwen2.5-Coder-7B by 8.9\%, while simultaneously improving functionality on HumanEval pass@1 by 2.1\%. Code is available at \href{https://github.com/viviable/MoC}{https://github.com/viviable/MoC}.
\end{abstract}

\section{Introduction}

Large language models (LLMs) have rapidly become useful tools for developers, demonstrating remarkable proficiency across a wide array of code generation tasks~\cite{chen2021evaluating,jiang2024survey}. Current LLMs excel at understanding complex programming concepts \cite{zheng2023codegeex}, generating syntactically correct and functionally relevant code \cite{hui2024qwen2,zhuo2025bigcodebench}, and even providing explanations, optimizations, and debugging assistance \cite{lewkowycz2022solving}.

\begin{figure}[t]
\centering
\includegraphics[width=\columnwidth]{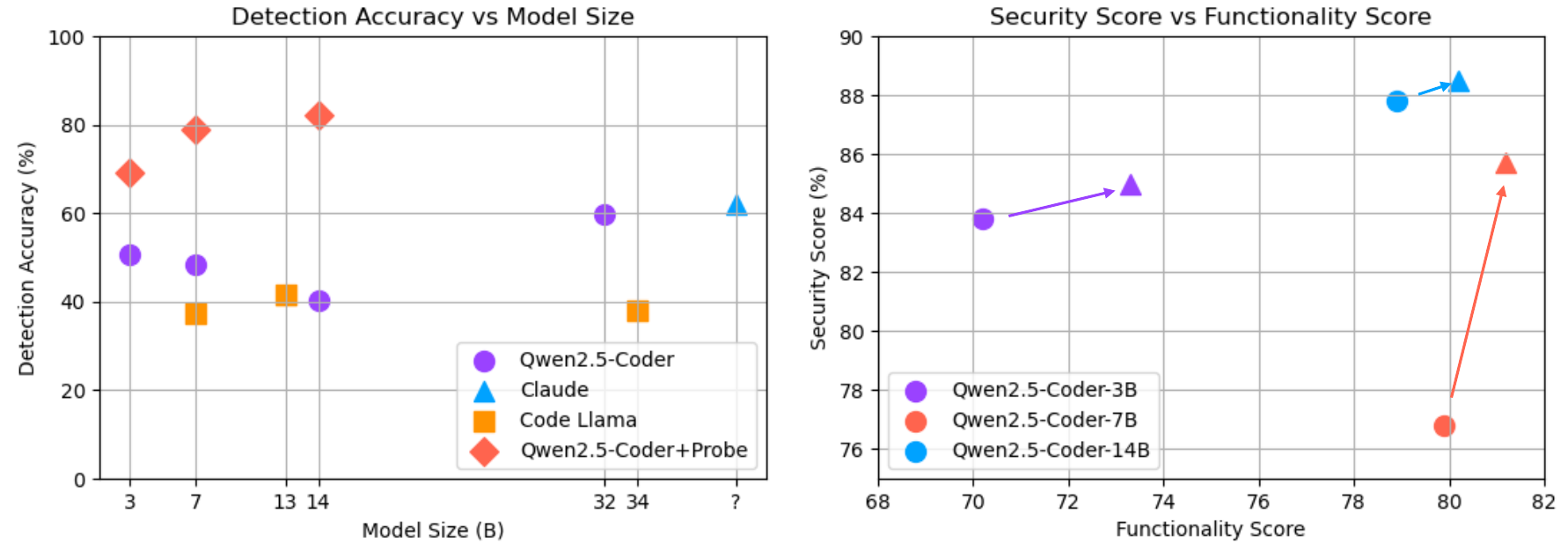} 
\caption{Left: The state-of-the-art code generation models cannot achieve high accuracy by purely prompting, while using probing can improve the accuracy. Right: Security and functional improvements by adding the mixture of corrections.}
\label{fig:few_prompts}
\end{figure}

Despite these advances, even state-of-the-art models exhibit significant limitations with identifying vulnerable code. 
Our empirical analysis (\autoref{fig:few_prompts}) on different sizes of CodeLlama~\cite{grattafiori2023code} and Qwen2.5-Coder~\cite{hui2024qwen2} reveals that traditional prompting techniques, including few-shot exemplars and detailed Common Weakness Enumeration (CWE) descriptions, result in accuracy comparable to random guessing (50\%). Surprisingly, increasing the model parameter count fails to reliably improve detection accuracy, suggesting a persistent gaps between increased coding capabilities and the closely related task of identifying and generating \emph{secure} code.

This motivates the question: \emph{do code-generating LLMs inherently lack the knowledge to differentiate between vulnerable and secure code, or is this knowledge simply not accessible via prompting?}
Using linear probing~\cite{alain2017understanding,zou2023representation}, we find that LLMs do indeed possess latent representations that distinguish secure from vulnerable code far more effectively than standard prompts. 
Thus, despite these models' apparent lack of proficiency at the task of identifying vulnerable code, it is possible to access models' precise learned knowledge about vulnerabilities to accurately perform identification during inference, without the need for more expensive methods involving fine-tuning~\cite{fu2022linevul}.

Building on this insight, we next investigate whether these latent representations can be leveraged during code generation. Specifically, we explore how to compute \emph{correction vectors}--derived directly from clusters,  linear probes, or through auxiliary neural networks--that encode vulnerability distinctions. These vectors, computed separately for individual CWEs, create a \emph{mixture} of precise linear corrections.

We integrate these guiding vectors into the model's token generation process, applying conditional corrections with a temporal decay to subtly adjust next-token probabilities based on vulnerability, as assessed from the linear probes. This method enables granular, controlled steering of generation away from vulnerable code while avoiding interference with generation unlikely to yield vulnerable code, and thus without sacrificing functionality. Importantly, we show that it is also possible to apply this process \emph{adversarially} to deliberately increase the likelihood of generating vulnerable code; this may be useful when training future models not to generate vulnerable code.

Our evaluation shows that this conditional steering not only significantly improves the security ratio (8.9\% on Qwen2.5-Coder), but also frequently enhances the functional correctness of the resulting code (2.1\% on HumanEval) (\autoref{fig:few_prompts}). 
Moreover, we observe that the guiding vectors often \emph{transfer} across models: vectores derived from one model can improve security in code generated by another model, such as the Qwen-2.5 variants. This transferability yields a computationally efficient way to harden models that are not well trained specifically on code data.

\section{Related Work}
\paragraph{LLM-assisted Vulnerability Detection} 
Vulnerability detection is a crucial task in the field of computer security. Its primary objective is to identify potential software security threats, thus reducing the risk of cyber-attacks. LLMs have been explored for vulnerability detection in source code using two main paradigms: fine-tuning and prompt engineering~\cite{ding2024vulnerability}. Fine-tuning approaches typically introduce a binary classification head on top of the LLM and jointly optimize all model parameters using labeled vulnerable and secure code examples \cite{du2024generalization}. This setup has been applied across various Transformer architectures, including encoder-only~\cite{kim2022vuldebert, sun2023assbert}, encoder-decoder~\cite{fu2022vulrepair}, and decoder-only models~\cite{zhou2024large}. Some methods \cite{yang2024security} also use Graph Neural Network backbone to extra features, and concatenate with LLM extracted features. Prompting-based methods~\cite{fu2023chatgpt} instead query powerful, often proprietary LLMs like GPT-4 using crafted natural language prompts. While these techniques have shown promising results on synthetic datasets~\cite{khare2023understanding}, their performance on real-world vulnerability detection tasks is mixed. More recent work has explored structured prompting strategies, such as variations of Chain-of-Thought (CoT) prompting~\cite{ullah2024llms}, and task-specific prompting frameworks targeting vulnerabilities like Use-Before-Initialization~\cite{li2024enhancing} and smart contract bugs~\cite{sun2024llm4vuln}.
Despite these advances, empirical studies have shown that both fine-tuning and prompting-based methods still struggle with vulnerability detection tasks~\cite{ding2024vulnerability}. 

In this work, we propose a new direction: instead of relying on extra finetuning or prompting strategies, we focus on representation engineering of trained LLMs to improve their internal understanding of secure and vulnerable code; we aim to enhance the latent representations used by the model to reason about code, enabling more robust vulnerability detection without introducing costly retraining or additional inference-time prompt engineering.

\paragraph{LLM-assisted Secure Code Generation}
Recent advancements in large language models (LLMs) have demonstrated significant potential in automating code generation tasks. However, the security of the code produced by these models remains a pressing concern, prompting a surge of research into security-aware techniques. Various approaches have been proposed to enhance the security of code generated by LLMs, focusing on both training-time and inference-time interventions. 

Techniques such as SafeCoder~\cite{heinstruction} and ProSec~\cite{xu2024prosec} use security-centric fine-tuning to improve security, utility, and alignment. 
APILOT~\cite{bai2024apilot} addresses the challenge of outdated or insecure API usage by implementing a mechanism that navigates LLMs to generate secure, version-aware code, thereby reducing potential security threats associated with deprecated APIs. INDICT~\cite{le2024indict} presents a multi-agent framework that employs internal dialogues between safety-driven and helpfulness-driven critics to iteratively refine code generation, enhancing both the security and functionality of the output.
CodeFavor~\cite{liu2024learning} proposes a code preference model trained on synthetic evolution data, including code commits and critiques, to predict whether a code snippet adheres to secure coding practices. While it does not directly generate code, it provides a mechanism to evaluate and prefer secure code snippets.

SVEN~\cite{he2023large} is closest to our work; it introduces a method that guides LLMs to generate secure or insecure code by learning a continuous prompt, without modifying the model's weights. This approach allows for controlled code generation based on specified security properties.
In contrast to SVEN, we compute a mixture of correction vectors, which are applied conditionally, leading to better control of the code generation. We compare in more detail with SVEN in section \ref{sec:exp}.

\paragraph{Steering \& Controlling LLM Generation}

Controllable generation refers to the ability to steer the outputs of large language models (LLMs) toward desired properties, such as stylistic attributes, factuality, safety, or personalization. A growing body of work has focused on developing techniques for controlling LLMs both at the input and internal representation levels \cite{liang2024controllabletextgenerationlarge}. A prominent strategy for understanding and influencing LLM behavior is probing, which involves training lightweight classifiers on the model's internal activations to extract human-interpretable features \cite{alain2017understanding}. Probing has been widely used to reveal latent knowledge in language models and, more recently, to guide and steer generation behavior by identifying representation subspaces associated with specific attributes. Recent advances in representation engineering go beyond passive probing, proposing direct interventions in the model’s latent space. These techniques identify semantically meaningful directions in activation space and apply steering vectors to modify model behavior without full retraining \cite{wehner2025taxonomy,zou2023representation}. Such approaches have been used for tasks like factuality correction, sentiment control, and personalized generation \cite{caopersonalized,wang2025adaptive}. Despite this progress, only a few studies~\cite{he2023large,pimparkhede2024doccgen} have explored the application of controllable generation techniques to code generation, where correctness, determinism, and alignment with developer intent are critical. In this paper, we propose a novel framework that applies probing and representation interventions to code generation models. Our method performs conditional interventions in the activation space, guided by the outputs of probes trained to detect semantic properties or vulnerabilities in the code.

\section{A Mixture of Linear Corrections}

\begin{figure}[t]
\centering
\includegraphics[width=\columnwidth]{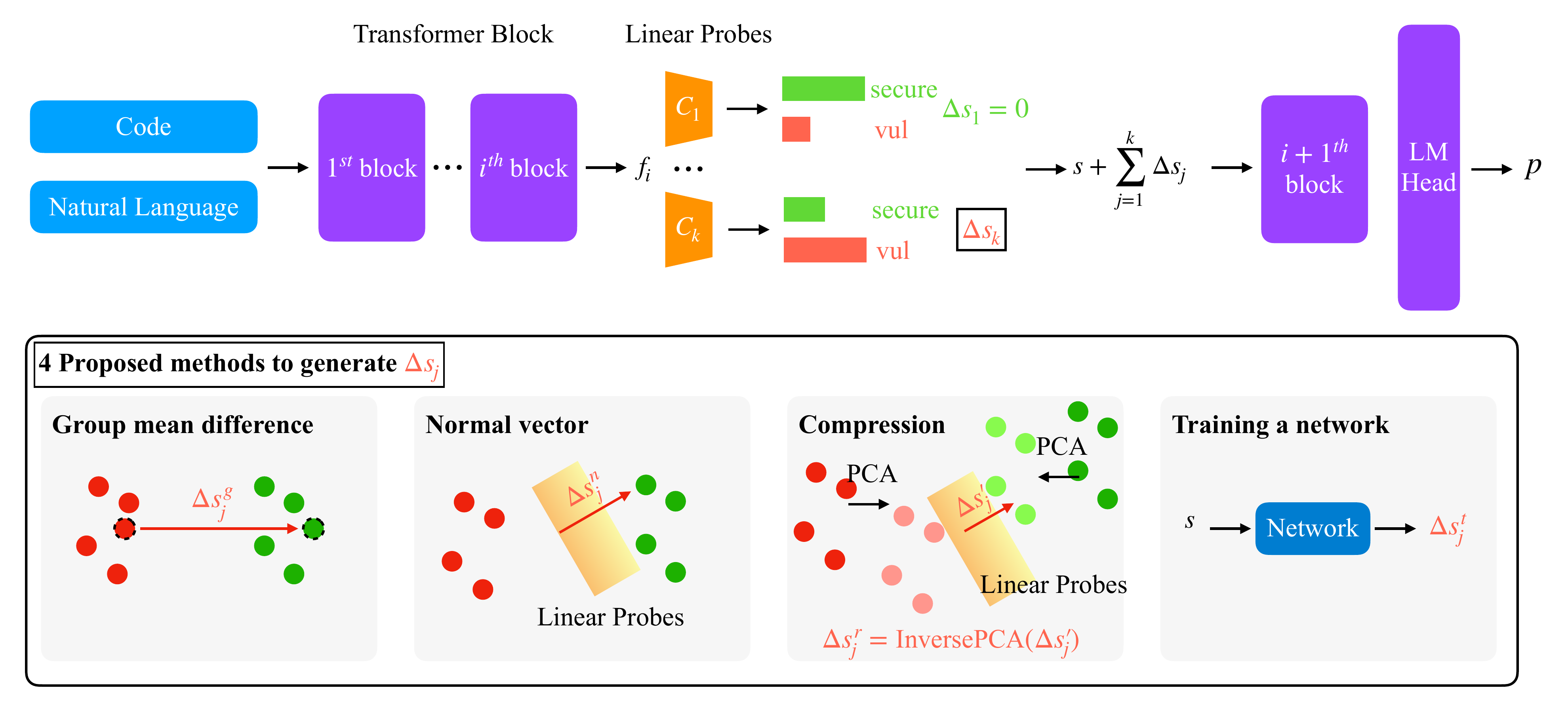} 
\caption{Mixture of corrections (MoC). There are four ways to obtain corrections for each vulnerability $j$. During inference, MoC applies correction $\Delta s_j^{c\in \{g,n,r,t\}}$ \textcolor[HTML]{FF644E}{if the hidden states are at risk of generating $j^{th}$ type of vulnerable code}, and \textcolor[HTML]{7ECF7E}{won't make correction if secure}.}
\label{fig:pipeline}
\end{figure}

Figure~\ref{fig:pipeline} gives an overview of our approach. We first train a set of linear probes for code vulnerability detection, as described in \autoref{sec:detection}, which we then use alongside one of four methods to obtain a set of linear corrections, as described in \autoref{sec:correction}. Finally, we use a mixture of the corrections, one for each vulnerability class, to generate secure code, as described in \autoref{sec:inference}.

\subsection{Vulnerable Code Detection Using Linear Probing} \label{sec:detection}

We are given a decoder model used for code generation, denoted $\mathcal{G}$, with transfomer blocks $L_0, \ldots L_n$. 
We denote the $d$-dimensional activations of the hidden state at the last token position of a block $L_i$ as $s_i$.
A \emph{probe} is a diagnostic tool that analyzes the information represented $s_i$, for a particular block $i$.
Given a dataset $\mathcal{D}$ of paired (secure, vulnerable) code samples $\{(x^+, x^-)\}$ and a vulnerability type $j$, we will write $\mathcal{D}_j$ to refer to the subset of $\mathcal{D}$ consisting only of vulnerability type $j$.

From the dataset, we use cross entropy loss to train a set of linear probes $c_0(\cdot), \cdots c_k(\cdot)$ against a binary label $v$ which denotes whether the features $s$ were produced by a vulnerable ($x^-$) or secure ($x^+$) sample (\autoref{eq:probe}).  
\begin{equation} \label{eq:probe}
    \mathcal{L}_{c} = \textrm{CE}(c(s), v) = \textrm{CE}(Ws + b, v)
\end{equation}
We perform this training on all blocks in the model, and identify the block $L^{*}$ with the smallest loss to take as the final probe for each vulnerability.

Our empirical investigation reveals that the activations within an LLM exhibit remarkable efficacy for vulnerability detection and the detection accuracy is good compared to previous finetuned or GNN-based detectors. This finding is also a proof that hidden states within LLMs encode richer information than the terminal outputs, as shown in \cite{azaria2023internal,chen2024states}, Notably, training these probes only requires minimal data with lightweight parameters.

\subsection{Controlling Vulnerability Generation with a Mixture of Corrections (MoC)}  \label{sec:correction}

The efficacy of vulnerability detection through representational probing demonstrates that the latent activations within transformer attention mechanisms encode substantive information pertaining to code security vulnerabilities. This observation suggests that these representations can be leveraged to guide code generation—-either to make code more secure, or to deliberately produce vulnerabilities. 
We propose a framework called \emph{Mixture of Corrections} (MoC) for accomplishing this, which computes a set (mixture) of correction vectors $\Delta s_j$ for each vulnerability class, which are subsequently combined with hidden states when generating code during inference.

We present four methods for computing correction vectors, both static, wherein $\Delta s_j$ is a function only of the vulnerability type $j$, as well as dynamic, where it is also conditioned on the decoder's current hidden states.
MoC is illustrated in \autoref{fig:pipeline}, and detailed in alg. \ref{alg: moc}.

\subsubsection{Static Correction Vectors}\label{sec:static}

\textbf{Difference of group mean}. The most direct and efficient way of measuring the correction is by computing the arithmetic differential between the centroid vectors of the respective class data samples, as shown in \autoref{eq:gm}.
\begin{equation} \label{eq:gm}
    \Delta s_j^{g} = \frac{1}{|s_j^+|}\sum_{\mathcal{D}_j} s_j^+ - \frac{1}{|s_j^-|}\sum_{\mathcal{D}_j} s_j^-
\end{equation}
In~\autoref{eq:gm}, $|\cdot|$ denotes the cardinality of the set of $j^{th}$ vulnerability.

\textbf{Normal vector of the decision boundary}. The linear probe established in the preceding section effectively partitions the feature space into two disjoint subspaces via a linear hyperplane that constitutes the decision boundary. Thus, another way of computing the requisite correction is to traverse orthogonally from the vulnerable class subspace to the non-vulnerable class subspace—specifically, the normal vector to the decision boundary hyperplane. The decision boundary is $(W_{1,:}-W_{0,:})x +b_1-b_0=0$, and the normal vector is characterized in \autoref{eq:nv}.
\begin{equation} \label{eq:nv}
    \Delta s_j^{n} = W_{1,:}-W_{0,:}
\end{equation}
In~\autoref{eq:nv}, $W \in R^{2\times d}$, and $W_{0,:}$ and $W_{1,:}$ is the first and second row of $W$.

\textbf{Reduced normal vector}. Direct utilization of LLM hidden states can exhibit susceptibility to overfitting phenomena and manifest training instability in probe training. Our assumption is that within the high-dimensional feature space, these features encode not only vulnerability-related information but also other types of information that can be considered as noise in code generation contexts. To mitigate these adverse effects, we use dimensionality reduction techniques, specifically principal component analysis (PCA), to derive a more robust correction vector \cite{zou2023representation}. Let $s_j'$ to denote the compressed version of $s_j$, and we train a linear probe $c_j'(x)=W'x+b'$ on the compressed vectors, then project the correction back to the original high-dimensional space, i.e.
$
    \Delta s_j^{r} = \textrm{PCAInverse} (c_j')
$,
where $W' \in R^{2\times d'}$, $d'$ is the reduced dimension. 

\subsubsection{Dynamic Correction Vectors} \label{sec:dynamic}
To condition the correction vectors additionally on the dynamic state of the model, for each vulnerability type we train a neural network $N(\cdot)$ that directly predicts the correction vector $\Delta s^t$. 
To train this network---given that the vulnerability dataset usually contains not only the paired data, but also the detailed line changes of the vulnerable lines of code---we add multiple aspects of supervision. Let $p_j$ denote the $\mathcal{G}$'s output probability corresponding to $s_j$ at the hidden space, and let $y^+$ denote the secure code labels in token space. Note that the $p_j$, $s_j$ and $y^+$ here are per-token supervision, and should be denoted as $p_{j,m}, p_{j,m+1}, \cdots, p_{j,m+n}$, where $m$ is the index for the start of the vulnerable code token, and $m+n$ is the index for the end of the vulnerable code token. We use the $p_j$ for simplicity. 

The training involves three loss terms, including mean square error, $\mathcal{L}_{mse} = \textrm{MSE}(s_j^-, s_j^+)$, cross entropy loss
$\mathcal{L}_{ce} = \textrm{CE}(p_j^-, y_j^+)$, and KL-divergence,
$\mathcal{L}_{KL} = \textrm{KL}(p_j^-, p_j^+)$. The final loss form is a combination of these supervisions,
$\mathcal{L} = \beta_1 \mathcal{L}_{mse}+\beta_2 \mathcal{L}_{ce} + \beta_3 \mathcal{L}_{KL}.$ 
The network then gives the correction, as
$
\Delta s_j^t = N(s_j)
$.

\subsection{Inference with corrections} \label{sec:inference}
After obtaining the mixture of corrections $\{\Delta s_j\}$, one for each vulnerability class, we apply the corrections during inference time, by adding the linear combination of corrections to the hidden states when generating every token. However, unlike the works \cite{bhattacharjee2024towards,rimsky2024steering} that directly add the vectors, i.e., $f = f + \Delta s$, we find it sub-optimal and add the following tricks.

\textbf{Decay}. During inference, as the generation becomes longer, the correction accumulates, which results in too much correction and the output generations can be less meaningful. To avoid such a large change in hidden states, we use a decay factor to gradually reduce the impact of the newly added correction during inference.
\begin{equation}
    \Delta s := \alpha(t)\cdot\Delta s,
\end{equation}
where $t$ is the number of tokens newly generated during inference, e.g. when generating the first token, $t=0$, and when generating the $k^{th}$ token, $t=k-1$. $\alpha(\cdot)$ is a negative exponential function. 

\textbf{Conditional correction}. When generating secure code, there's no need to modify the hidden states and add the correction vectors. The corrections are only applied when the hidden states are at risk of generating vulnerable codes. Thus, we apply a conditional correction, as in the \autoref{fig:pipeline}, we first use the previously obtained linear probes to detect if the current hidden states are at risk of generating vulnerable codes of vulnerable type $j$, and then apply the corresponding correction $\Delta s_j$ only if the hidden states are vulnerable. If the current hidden states shows multiple different vulnerabilities, then the corrections are added as a linear combination as shown in \autoref{eq:lc}.
\begin{equation} \label{eq:lc}
    s += 
\begin{cases}
    \Delta s_j,  & \textrm{if } \textrm{argmax}(c_j(s)) = 0   \\
    0,           & \textrm{otherwise}
\end{cases}
\end{equation}

We present the overall MoC algorithm in alg. \ref{alg: moc}. MoC first trains the light-weight linear probes, and then obtains corrections for each type of vulnerability. During inference, MoC applies these corrections if the hidden states are at risk of generating vulnerable code, as measured by the linear probes.

\SetKwInput{KwInput}{Input} 
\SetKwInput{KwOutput}{Output} 
\SetKwRepeat{Do}{do}{while}
\SetKwIF{If}{ElseIf}{Else}{if}{then}{else if}{else}{}
\begin{algorithm}[t] \label{alg: moc}
\SetAlgoLined
\KwInput{(1) A code generation LLM $\mathcal{G}$, and its $i^{th}$ transformer blocks $L_i$; (2) A dataset of paired vulnerable and secure data  $\mathcal{D} = \{\mathcal{D}_j\}$. }
\KwOutput{A secure code generation $x$}
\tcp{Training stage}
\ForEach{$j\in \{0,\ldots,k\}$}{
    Train the linear probe $c_j$ according to \autoref{sec:detection}. \\
    Obtain the correction vector $\Delta s_j^{c \in \{g,n,r,t\}}$ according to \autoref{sec:static} \& \autoref{sec:dynamic}\\}
\tcp{Inference stage}
\ForEach{token $x_{t+1}$}{
    $s = L^{*}(x_{1:t})$  \tcp*{Get the hidden states} 
    \ForEach{$j\in \{0,\ldots,k\}$}{
        \uIf{($argmax(c_j(s)) = 0)$}{ 
            $s:=s + \alpha(t)\cdot\Delta s_j^m$  \tcp*{add correction if vulnerable}
        }
    }
}
return $x$
\caption{Mixture of Corrections}
\label{alg: moc}
\end{algorithm}

\section{Experiments}
\label{sec:exp}
In this section, we first present the evaluation of the trained probe's efficacy in identifying vulnerable code, then we investigate whether the mixture of corrections improves, or adversarially, decreases security, in code generation, and the transferability of the corrections across models. 

\subsection{Vulnerable Code Detection} \label{sec: exp_detection}

\textbf{Dataset}. Following SVEN \cite{he2023large}, which contributes a high-quality pairwise code dataset of 9 different CWEs, we use this dataset as our training and evaluation set. In each vulnerability class, we random sample a train set and a subset. Due to the imbalance of the dataset across different types of vulnerabilities, we keep the evaluation set the same size, and the train set might be of different sizes. Notably, the vulnerable and secure data are balanced in our settings.

\textbf{Evaluation Metric}. Accuracy $\textrm{Acc}_v$ (\%) and $\textrm{Acc}_s$ (\%) is the accuracy of the vulnerable code and secure code respectively. For training, $\textrm{Acc}$ (\%) and F1 (range from 0 to 1) are the accuracy and F1-score on the evaluation set. 

\textbf{Linear Probe Details.} For each vulnerability, the probe is trained on around 50 to 150 data points due to the imbalance of different types of vulnerabilities. The training epoch is from 50 to 200, with a batch size of 64, learning rate $5e-4$, SGD optimizer with a momentum and weight decay.

\textbf{Training Details.} One of the proposed correction methods requires training of another network, and the network structure is a three-layer multi-layer perception (MLP),  with GeLU activation, layer normalization, and dropout layer. The learning rate is $1e-3$, with an Adam optimizer. To construct the training pair $f_i^+$ and $f_i^-$, we also use the `line changes' information in each pair of the vulnerable-secure data for detailed supervision. We save the hidden states tensors before training so that training is done on single GPU even for 14B models.

\textbf{RQ1: Can LLMs detect vulnerable code by direct prompting?}  As evidenced in \autoref{tab: direct_prompt}, by directly prompting the code generation LLMs, the accuracy is suboptimal. Notably, the prompt includes both few-shot examples sampled from the same dataset (two positive examples and two negative examples), and a description of the specific vulnerability (e.g. CWE-022). We list per-vulnerability experimental results in \autoref{append: direct_prompt}. We can draw the following conclusions: (1) In general, the current code-related LLMs, including Qwen2.5-Coder series and CodeLlama series, and closed-source model Claude lack the ability to detect code vulnerabilities by prompting. Possible reasons are that the vulnerabilities are less focused on and that these models are not specifically trained on vulnerability code data. (2) There is no clear relation between the model size and their vulnerable detection capacity, though the 32B or 34B models show a small performance improvement compared to smaller models. (3) QC-7B, 14B and 32B, CL-34B models tend to predict the code secure. For the QC-7B, 14B, and CL series, the accuracy is no better than a random guess.
\begin{figure}[htbp]
  \centering
  \begin{minipage}[c]{0.43\textwidth}
    \centering
    \captionof{table}{Accuracy of vulnerable code detection by direct prompting the LLM. Invalid means the output of the LLM doesn't follow the format or the output does not include any answers. QC is in short for Qwen2.5-Coder, and CL for Code Llama.} 
    \label{tab: direct_prompt}
    \begin{tabular}{lccc}
      \toprule
                   & $\textrm{Acc}_v$  &$\textrm{Acc}_s$ & Invalid  \\ \midrule
      QC-3B        & 51                & 51              & 0        \\
      QC-7B        & 23                & 74              & 3        \\
      QC-14B       & 25                & 55              & 27       \\
      QC-32B       & 30                & 81              & 0        \\ \midrule
      CL-7B        & 64                & 10              & 29       \\
      CL-13B       & 44                & 43              & 14       \\
      CL-34B       & 21                & 59              & 27       \\ \midrule
      Claude       & 63                & 43              & 0        \\ \bottomrule          
    \end{tabular}
  \end{minipage}
  \hfill
  \begin{minipage}[c]{0.53\textwidth}
    \begin{tcolorbox}[
      colback=gray!10,
      colframe=gray!50,
      width=\textwidth, 
      height=7cm,
      title=Prompts
    ]
    // CWE description
    
    CWE-022, commonly called "Path Traversal," is a vulnerability when an application fails to appropriately limit … 
    
    // Few-shot examples
    
    For example,  'code1 …' is vulnerable, while 'code2 …' is not vulnerable. 'code3 …' is vulnerable, while 'code4 …' is not vulnerable. 
    
    // Prompt
    
    Is the subsequent code susceptible to the specified vulnerability?
    
    // Test code
    
    code …
    
    Answer the question with simply yes or no.
    \end{tcolorbox}
  \end{minipage}
\end{figure}

\begin{table}[!h]
\centering
\caption{Performance on code vulnerability detection.}
\begin{tabular}{lcccccc}
\toprule
\multirow{2}{*}{\textbf{Method}} & \multicolumn{2}{c}{\textbf{QC-3B}} & \multicolumn{2}{c}{\textbf{QC-7B}} & \multicolumn{2}{c}{\textbf{QC-14B}} \\
\cmidrule(lr){2-3} \cmidrule(lr){4-5} \cmidrule(lr){6-7}
 & \textbf{Acc} & \textbf{F1} & \textbf{Acc} & \textbf{F1} & \textbf{Acc} & \textbf{F1} \\
\midrule
Prompting                  & 51 & 0.56 & 49 & 0.53 & 40 & 0.37 \\
Linear Probe W/O Few-shot  & 66 & 0.65 & 68 & 0.66 & 75 & 0.79 \\
Linear Probe               & 69 & 0.63 & \textbf{79} & \textbf{0.76} & \textbf{82} & \textbf{0.85} \\
Linear Probe PCA           & \textbf{72} & \textbf{0.68} & 76 & 0.74 & 78 & 0.80 \\
MLP Probe                  & \textbf{72} & 0.66 & 77 & 0.75 & 80 & 0.80 \\
\bottomrule
\end{tabular}
\label{tab:detection}
\end{table}

\textbf{RQ2: Can hidden states within LLMs help detect vulnerable code?} In \autoref{tab:detection}, `Prompting' means no probe training, and just prompting by few-shot and descriptions as in RQ1. `Linear Probe W/O Few-shot' refers to, when getting hidden states $f$ from $L_i$ in $\mathcal{G}$, the input only includes the code without few-shot examples. The other probes' input all includes few-shot examples. `Linear Probe' and `Linear Probe PCA' contains a linear layer with a weight matrix $W$ and bias $b$, the difference is without PCA $W \in R^{2 \times d}$, where $d=3168$ in this cases, with PCA, $W' \in R^{2 \times d'}$ and $d'$ is a number between 50 to 100. `MLP probe' contains 2 or 3 multi-layer perceptron layers, each layer includes a linear layer, a ReLU activation function, a layer norm, and a dropout layer. 

From \autoref{tab:detection}, we can draw the following conclusion. (1) Overall, probing methods can detect vulnerable code, showing that hidden states within LLMs actually contain vulnerability-related information. (2) Using few-shot examples in the text prompt improves the vulnerability detection, showing that the prompting techniques help with the hidden states probing. (3) MLP probes, even with more parameters, don't show a clear improvement compared to linear probes. This may be due to the simplicity of the task: it is a classification task and linear probes are enough to distinguish the secure and vulnerable classes. (4) The performance shows a relation with the LLM scale, as the LLM becomes larger, the performance of the probe gets higher.

\subsection{Secure Code Generation}

\textbf{Evaluation}. We evaluate the code security using GitHub CodeQL \cite{codeql}, which is an open-source code security analyzer that can detect different vulnerabilities based on the custom queries. We report the security rate SR (\%). $\text{SR}_h$ means hardening the security (the higher the better), while $\text{SR}_w$ means weakening the security (the lower the better). The generated code is considered secure only if it doesn't contains any main CWEs based on CodeQL. Note that we test the proposed methods on the SVEN test set, which is different from the evaluation set in \autoref{sec: exp_detection}. For code functionality, we test the pass@1 on HumanEval.

\textbf{RQ3: Can the mixture of corrections help in secure code generation?} In \autoref{tab: gen}, `Base Model' means applying no corrections. `SVEN' \cite{he2023large} is a method that trains prefix soft embeddings and concatenates the embeddings to the LLM during inference. However, on the 3B model, the training loss doesn't decrease, so we choose not to report the results. Then the four correction methods refer to $\Delta s_j^g, \Delta s_j^n, \Delta s_j^r, \Delta s_j^t$ respectively. In the security hardening cases, the conditional generation is utilized, while in the security weakening cases, since the aim is to modify the secure hidden states to insecure ones, and thus it is not conditional, we add the sum of the negative corrections to it, i.e. $\Delta s=\sum_{j=1}^{k}-\alpha(t)s_j$.

We can draw the conclusion that: (1) Generally, applying MoC can improve not only the security but also the functionality of the code. (2) On Qwen-2.5-Coder-7B, the dynamic NN-based method outperforms others. (3) There are some cases when the weakening cases do not actually bring out more vulnerable codes. One possible guess is that, since we use the sum of all the correction vectors, they may suffer a bit by canceling out on some critical directions. (4) In most cases, the functionality of the LLMs is not affected and even shows some improvements.

\begin{table}[]
\caption{Performance on code generation. $\textrm{SR}_h$ ($\uparrow$)(\%) and $\textrm{SR}_w$ ($\downarrow$)(\%) denote the security ratio when applying hardening and weakening. HE denotes HumanEval pass@1. } \label{tab: gen}
\begin{center}
\begin{tabular}{cccccccccc}
\toprule
            & \multicolumn{3}{c}{QC-3B}                                                         & \multicolumn{3}{c}{QC-7B}                                                        & \multicolumn{3}{c}{QC-14B}                                                         \\ \midrule
            & $\textrm{SR}_h$      & $\textrm{SR}_w$       & HE                & $\textrm{SR}_h$     & $\textrm{SR}_w$       & HE                & $\textrm{SR}_h$       & $\textrm{SR}_w$    & HE                \\ \midrule
Base Model   & 83.8                     & 83.8                     & 70.2                     & 76.8                     & 76.8                     & 79.9                     & 87.8                     & 87.8                     & 78.9                     \\
SVEN        &         -                 &                -          &   -                       &              65.0            &   54.0                       &         75.3                 &    69.7                      &        65.4                  &               75.0           \\ \midrule
Group Mean Diff     & {84.7}   & {78.8} & {73.9} & {84.0}   & {75.5} & {81.4} & {88.5}   & {87.1} & {80.2} \\
Normal Vector      & 83.3 &        81.0                  & {74.5}   & {84.3} &                 {80.4}         & {82.0}   & 87.5 &   87.2 & {78.9}   \\
Normal Vector PCA &     85.0                     & 82.4 & 73.3 & 82.9 & 78.9 & 82.0 & 88.3 & 82.5 & 78.3 \\
Dynamic NN-based    & 84.9   & 82.1 & 70.8 & {85.7}   & {75.9} & {81.2} &  88.0  & 87.1 & 82.0 \\ \bottomrule
\end{tabular}
\end{center}
\end{table}

\textbf{RQ4: Probes at which attention blocks are the best?} We train the probe on different attention blocks, and test their effect on the code generation. As in \autoref{fig:layer}, the last attention block shows the best performance.

\textbf{Ablation Study}. We conduct two versions for PCA corrections. One version is to first obtain both the decision boundary and compute the normal vector to the decision boundary in the compressed space, and then project the normal vector back to the high-dimensional space, as follows:
\begin{equation} \label{eq: pca0}
    \Delta s_j^{r} = \textrm{PCAInverse} (W'_{1,:}-W'_{0,:}) = (W'_{1,:}-W'_{0,:})V + M,
\end{equation}
where $V$ are the principal components and $M$ are the mean of the vectors.
Another is to first project the weighting matrix back and then calculate the normal vector, as follows:
\begin{equation} \label{eq: pca1}
    \Delta s_j^{r} = \textrm{PCAInverse} (W')_{1,:}-\textrm{PCAInverse} (W')_{0,:} = (W'V)_{1,:}-(W'V)_{0,:} + M_{1,:}-M_{0,:},
\end{equation}
note that $W \neq W'V$. We tried both, as in \autoref{tab: pca}, the first PCA version fails to generate reasonable outputs, while the second PCA implementation can bring improvements, suggesting that the hidden states space within LLMs is elaborate.

\begin{minipage}[t]{0.45\textwidth}
\begin{table}[H]
\centering
\caption{Ablation study on how to obtain PCA correction.} \label{tab: pca}
\begin{tabular}{cccc}
\toprule
            & $\textrm{SR}_h$      & $\textrm{SR}_w$       & HE                \\ \midrule
Base Model   & 76.8 & 76.8 & 79.9                     \\
   $\Delta s_j^{r}$ in \autoref{eq: pca0}  &       6.3                   &   4.6                       &       19.8                 \\ 
 $\Delta s_j^{r}$ in \autoref{eq: pca1}   &  82.9  &{78.9}  & {82.0}  \\ \bottomrule
\end{tabular}
\end{table}
\end{minipage}
\hfill
\begin{minipage}[t]{0.45\textwidth}
\begin{table}[H]
\centering
\caption{Ablation study on conditional correction and decay.} \label{tab: ablation}
\begin{tabular}{ccc}
\toprule
            & $\textrm{SR}_h$      &    HE                \\ \midrule
Base Model   & 76.8 &  79.9                     \\
  Normal Vector W/O Condition  &     81.7 &     77.0                \\ 
Normal Vector W/O Decay  & 85.8    &  69.6  \\
Normal Vector   &  84.3   & 82.0 \\ \bottomrule
\end{tabular}
\end{table}
\end{minipage}

Ablation in \autoref{tab: ablation} is conducted on QC-7B model. (1) Adding conditions improves both the secure ratio and the functionality. (2) Though adding decay results in an improvement on secure ratio, it affects the functionality significantly.

\textbf{RQ5: Can the corrections learned for one model transfer to another?} We try to apply the corrections learned from one model and apply them on another model. As in \autoref{tab: transfer}, the corrections are trained on Qwen2.5-Coder model and implemented on the Qwen2.5-Instruct model, where they share the same hidden dimension and the same model structure. We find that it shows some level of transferability in 3B and 7B models, but not on larger model. However, the functionality of the model based on transferred corrections are harmed on larger models.

\begin{minipage}[t]{0.45\textwidth}
\begin{table}[H]
\centering
\caption{Transferability across models.  We use QI in short for Qwen2.5-Instruct and QC for Qwen2.5-Coder. HE is short for HumanEval. The corrections are Normal Vector obtained from QC models.} \label{tab: transfer}
\begin{tabular}{ccccc}
\toprule
            &   Corrections    & $\textrm{SR}_h$      &  $\textrm{SR}_w$  & HE                \\ \midrule
QI-7B   &  & 76.8  &    76.8       &  69.6         \\
  QI-7B   &  QC-7B & 77.1   &    76.3 & 65.8   \\ 
 QI-3B &    & 75.3 & 75.3 &  54.0 \\ 
 QI-3B  &  QC-3B &   78.0 & 71.2 & 54.7 \\
  QI-14B &    & 63.6 & 63.6 & 74.5 \\ 
 QI-14B    &  QC-14B &   58.2 & 53.5 &  72.7\\\bottomrule
\end{tabular}
\end{table}
\end{minipage}
\hfill
\begin{minipage}[t]{0.45\textwidth}
\centering
\begin{figure}[H]
\centering
\includegraphics[width=\columnwidth]{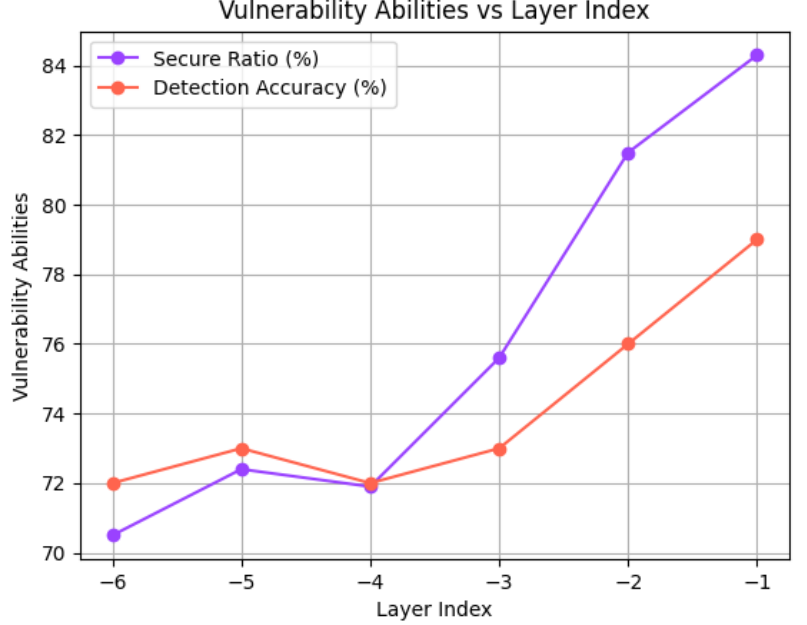} 
\caption{Ablations on $i^{th}$ attention blocks.}
\label{fig:layer}
\end{figure}
    
\end{minipage}

\textbf{RQ6: Why can the MoC gain the improvement on functionality for free?} As in \autoref{tab: gen}, we observe a consistent improvement on the functionality score except for the 14B model. A possible reason for this is that the more buggy codes have a higher possibility of also being vulnerable code \cite{morrison2015challenges}, and there are overlaps between bug-prone code and vulnerabilities \cite{camilo2015bugs}.

\section{Conclusion}
Our investigation reveals that code generation LLMs encode vulnerability-discriminative information in their hidden representations, accessible through lightweight linear probes. We leveraged this insight to develop a Mixture of Linear Corrections (MoC) framework that conditionally applies guiding vectors during inference to enhance code security. Experimental results show our method effectively improves both security ratios and functional correctness across multiple model sizes and demonstrates transferability between models. This work provides a computationally efficient approach to secure code generation without requiring costly retraining or extensive prompt engineering, opening new avenues for representation-based security interventions in generative AI.

\textbf{Limitations.} Currently we use every probe in every token generation, which consumes more time than the base model. Further work can implement all probes in parallel to accelerate. Moreover, the MoC can only address known code vulnerabilities, further research can focus on using hidden states to explore undiscovered vulnerabiliies.


\medskip
\small
\bibliographystyle{unsrt}
\bibliography{ref}

\newpage
\appendix

\section{Appendix}
\subsection{Detailed experimental results on direct prompting for vulnerability detection.} \label{append: direct_prompt}

Here we show the detailed experimental results for each CWEs for each model. The results are from the \autoref{append: qc3} to \autoref{append: cl34}, we find that: 
\begin{enumerate}
    \item Difference CWE types shows every different trends, for example, the CWE-125, CWE-190, CWE-416, CWE-476, CWE-787 contains mostly codes in language c, and Qwen-25-Coder series tend to think they are safe, as in \autoref{append: qc7}, \autoref{append: qc14} and \autoref{append: qc32}. And the CodeLlama series tend to regard the CWE-416, CWE-476 and CWE-787 as safe, as in \autoref{append: cl13} and \autoref{append: cl34}.
    \item Overall, QwenCoder series are more recently developed and shows better abilities than CodeLlama series. And overall, the QC series show a better instruction following ability than CL series, as the Invalid rates are lower.
\end{enumerate}

\begin{table}[ht]
\centering
\caption{Accuracy of vulnerable code detection by direct prompting the LLM. $\textrm{Invalid}_v$ and $\textrm{Invalid}_s$ mean the output of the LLM doesn't follow the format or the output does not include any answers.} \label{append: qc3}
\begin{tabular}{ccccc}
\toprule
\multicolumn{5}{c}{QwenCoder-3B}                                                                                                                       \\ \midrule
\multicolumn{1}{l}{Vul-type} & \multicolumn{1}{l}{$\textrm{Acc}_v$} & \multicolumn{1}{l}{$\textrm{Acc}_s$} & \multicolumn{1}{l}{$\textrm{Invalid}_v$} & \multicolumn{1}{l}{$\textrm{Invalid}_s$} \\ \midrule
22                           & 49                         & 49                         & 0                             & 0                             \\
78                           & 51                         & 43                         & 1                             & 0                             \\
79                           & 35                         & 58                         & 0                             & 0                             \\
89                           & 47                         & 61                         & 0                             & 1                             \\
125                          & 52                         & 45                         & 1                             & 0                             \\
190                          & 67                         & 53                         & 0                             & 3                             \\
416                          & 45                         & 62                         & 0                             & 0                             \\
476                          & 59                         & 44                         & 1                             & 1                             \\
787                          & 50                         & 42                         & 0                             & 0                             \\ \midrule
Ave      & 51                & 51               & 0                  & 1   \\
\bottomrule
\end{tabular}
\end{table}

\begin{table}[ht]
\centering
\caption{Accuracy of vulnerable code detection by direct prompting the LLM. $\textrm{Invalid}_v$ and $\textrm{Invalid}_s$ mean the output of the LLM doesn't follow the format or the output does not include any answers.} \label{append: qc7}
\begin{tabular}{ccccc}
\toprule
\multicolumn{5}{c}{QwenCoder-7B}                            \\ \midrule
Vul-type & $\textrm{Acc}_v$      & $\textrm{Acc}_s$ & $\textrm{Invalid}_v$   & $\textrm{Invalid}_s$   \\ \midrule
22       & 43          & 53     & 0           & 0           \\
78       & 66          & 72     & 0           & 0           \\
79       & 44          & 42     & 5           & 5           \\
89       & 1           & 69     & 0           & 1           \\
125      & 12          & 77     & 13          & 9           \\
190      & 8           & 100    & 0           & 0           \\
416      & 2           & 96     & 0           & 0           \\
476      & 13          & 84     & 4           & 3           \\
787      & 15          & 73     & 3           & 2           \\ \midrule
Ave      & 23 & 74     & 3 & 2 \\
\bottomrule
\end{tabular}
\end{table}

\begin{table}[ht]
\centering
\caption{Accuracy of vulnerable code detection by direct prompting the LLM. $\textrm{Invalid}_v$ and $\textrm{Invalid}_s$ mean the output of the LLM doesn't follow the format or the output does not include any answers.} \label{append: qc14}
\begin{tabular}{ccccc}
\toprule
\multicolumn{5}{c}{QwenCoder-14B}                                                                                                  \\ \midrule
Vul-type & \multicolumn{1}{c}{$\textrm{Acc}_v$} & \multicolumn{1}{c}{$\textrm{Acc}_s$} & \multicolumn{1}{c}{$\textrm{Invalid}_v$} & \multicolumn{1}{c}{$\textrm{Invalid}_s$} \\ \midrule
22       & 2                          & 31                         & 69                            & 69                            \\
78       & 25                         & 62                         & 12                            & 20                            \\
79       & 28                         & 72                         & 14                            & 14                            \\
89       & 13                         & 8                          & 78                            & 91                            \\
125      & 31                         & 64                         & 13                            & 14                            \\
190      & 36                         & 47                         & 19                            & 11                            \\
416      & 27                         & 71                         & 16                            & 15                            \\
476      & 32                         & 74                         & 3                             & 1                             \\
787      & 35                         & 63                         & 23                            & 6                             \\ \midrule
Ave      & 25                & 55                & 27                   & 27                  \\ \bottomrule
\end{tabular}
\end{table}

\begin{table}[ht]
\centering
\caption{Accuracy of vulnerable code detection by direct prompting the LLM. $\textrm{Invalid}_v$ and $\textrm{Invalid}_s$ mean the output of the LLM doesn't follow the format or the output does not include any answers.} \label{append: qc32}
\begin{tabular}{ccccc}
\toprule
\multicolumn{5}{c}{QwenCoder-32B}                                                                                                  \\ \midrule
Vul-type & \multicolumn{1}{c}{$\textrm{Acc}_v$} & \multicolumn{1}{c}{$\textrm{Acc}_s$} & \multicolumn{1}{c}{$\textrm{Invalid}_v$} & \multicolumn{1}{c}{$\textrm{Invalid}_s$} \\ \midrule
22       & 43                         & 45                         & 0                             & 0                             \\
78       & 69                         & 71                         & 0                             & 0                             \\
79       & 56                         & 44                         & 2                             & 2                             \\
89       & 99                         & 68                         & 0                             & 1                             \\
125      & 2                          & 100                        & 0                             & 0                             \\
190      & 0                          & 100                        & 0                             & 0                             \\
416      & 0                          & 100                        & 0                             & 0                             \\
476      & 0                          & 99                         & 0                             & 1                             \\
787      & 2                          & 100                        & 0                             & 0                             \\ \midrule
Ave      & 30                & 81                & 0                  & 0                 \\ \bottomrule
\end{tabular}
\end{table}

\begin{table}[ht]
\centering
\caption{Accuracy of vulnerable code detection by direct prompting the LLM. $\textrm{Invalid}_v$ and $\textrm{Invalid}_s$ mean the output of the LLM doesn't follow the format or the output does not include any answers.} \label{append: cl7}
\begin{tabular}{ccccc}
\toprule
\multicolumn{5}{c}{CodeLlama-7B}                                                                                                        \\ \midrule
Vul-type & $\textrm{Acc}_v$                          & $\textrm{Acc}_s$                 & $\textrm{Invalid}_v$                       & $\textrm{Invalid}_s$                       \\ \midrule
22       & 63                              & 0                      & 41                              & 37                              \\
78       & 49                              & 8                      & 48                              & 48                              \\
79       & 44                              & 9                      & 49                              & 51                              \\
89       & 0                               & 0                      & 100                             & 100                             \\
125      & 89                              & 11                     & 2                               & 2                               \\
190      & 86                              & 19                     & 3                               & 3                               \\
416      & 89                              & 9                      & 4                               & 2                               \\
476      & 59                              & 32                     & 12                              & 12                              \\
787      & 100                             & 2                      & 0                               & 0                               \\ \midrule
Ave      & 64 & 10 & 29 & 28 \\ \bottomrule
\end{tabular}
\end{table}

\begin{table}[]
\centering
\caption{Accuracy of vulnerable code detection by direct prompting the LLM. $\textrm{Invalid}_v$ and $\textrm{Invalid}_s$ mean the output of the LLM doesn't follow the format or the output does not include any answers.} \label{append: cl13}
\begin{tabular}{ccccc}
\toprule
\multicolumn{5}{c}{CodeLlama-13B}                                                                                                  \\ \midrule
Vul-type & \multicolumn{1}{c}{$\textrm{Acc}_v$} & \multicolumn{1}{c}{$\textrm{Acc}_s$} & \multicolumn{1}{c}{$\textrm{Invalid}_v$} & \multicolumn{1}{c}{$\textrm{Invalid}_s$} \\ \midrule
22       & 86                         & 14                         & 4                             & 4                             \\
78       & 63                         & 29                         & 10                            & 10                            \\
79       & 47                         & 42                         & 16                            & 21                            \\
89       & 65                         & 13                         & 26                            & 21                            \\
125      & 20                         & 18                         & 63                            & 62                            \\
190      & 83                         & 19                         & 0                             & 0                             \\
416      & 5                          & 96                         & 2                             & 0                             \\
476      & 8      & 72       & 2         & 3          \\
787      & 20       & 89       & 0          & 1         \\ \midrule
Ave      & 44              & 43                         & 14                  & 13     \\ \bottomrule             
\end{tabular}
\end{table}

\begin{table}[]
\centering
\caption{Accuracy of vulnerable code detection by direct prompting the LLM. $\textrm{Invalid}_v$ and $\textrm{Invalid}_s$ mean the output of the LLM doesn't follow the format or the output does not include any answers.} \label{append: cl34}
\begin{tabular}{crrrr}
\toprule
\multicolumn{5}{c}{CodeLlama-34B}                                                                                                  \\ \midrule
Vul-type & \multicolumn{1}{c}{$\textrm{Acc}_v$} & \multicolumn{1}{c}{$\textrm{Acc}_s$} & \multicolumn{1}{c}{$\textrm{Invalid}_v$} & \multicolumn{1}{c}{$\textrm{Invalid}_s$} \\ \midrule
22       & 61                         & 49                         & 8                             & 6                             \\
78       & 55                         & 37                         & 17                            & 15                            \\
79       & 7                          & 2                          & 90                            & 90                            \\
89       & 53                         & 60                         & 2                             & 3                             \\
125      & 1                          & 54                         & 49                            & 46                            \\
190      & 0                          & 89                         & 19                            & 17                            \\
416      & 2                          & 85                         & 15                            & 15                            \\
476      & 0                          & 81                         & 21                            & 22                            \\
787      & 10                         & 75                         & 23                            & 19                            \\ \midrule
Ave      & 21                         & 59                & 27                 & 26                 \\ \bottomrule
\end{tabular}
\end{table}

\subsection{Detailed experimental results on probing for vulnerability detection.} 
Here we shows more results about detailed per-CWE results on detection when training a linear probe on the last attention block. We can draw the conclusion:
\begin{enumerate}
    \item Overall, the non-PCA probe in \autoref{append: detection-linear} shows better results than PCA reduced probes in \autoref{append: detection-linear_pca}. Possible reasons are that the PCA reduced too much information that may be essential for vulnerability detection.
    \item Overall, the CWE-022, CWE-078, CWE-079, and CWE-089 are mostly based on python language. And these shows a higher accuracy than other CWEs, especially on QC-14B and 7B models.
    \item The CWE-125 and CWE-476 are kind of hard to detect, especially, as the model gets larger, the accuracy on these two CWE-types are not getting higher, which indicates that their vulnerable features are harder to extract.
\end{enumerate}
\begin{table}[ht]
\centering
\caption{Performance on code vulnerability detection.} \label{append: detection-linear}
\begin{tabular}{ccccccc}
\toprule
         & \multicolumn{2}{l}{QC-3B}                        & \multicolumn{2}{c}{QC-7B}                                       & \multicolumn{2}{c}{QC-14B}                                      \\ \midrule
Vul-type & {Acc} & {F1} & {Acc} & {F1} & {Acc} & {F1} \\ \midrule
22       & 0.8                              & 0.83                 & 0.9                                & 0.89                              & 0.7                                & 0.73                              \\
78       & 0.7                              & 0.67                 & 0.9                                & 0.91                              & 0.9                                & 0.91                              \\
79       & 0.7                              & 0.67                 & 0.9                                & 0.89                              & 0.9                                & 0.89                              \\
89       & 0.8                              & 0.75                 & 0.8                                & 0.75                              & 1                                  & 1                                 \\
125      & 0.7                              & 0.73                 & 0.7                                & 0.67                              & 0.6                                & 0.67                              \\
190      & 0.7                              & 0.57                 & 0.6                                & 0.67                              & 0.8                                & 0.8                               \\
416      & 0.8                              & 0.75                 & 0.7                                & 0.57                              & 0.8                                & 0.8                               \\
476      & 0.6                              & 0.6                  & 0.6                                & 0.67                              & 0.6                                & 0.67                              \\
787      & 0.7                              & 0.57                 & 0.7                                & 0.67                              & 0.7                                & 0.73                              \\ \midrule
Ave      & 0.72                     & 0.68         & 0.76                       & 0.74                      & 0.78                       & 0.8   \\ \bottomrule                           
\end{tabular}
\end{table}

\begin{table}[]
\centering
\caption{Performance on code vulnerability detection. PCA is applied to the hidden states. } \label{append: detection-linear_pca}
\begin{tabular}{ccccccc}
\toprule
         & \multicolumn{2}{c}{QC-3B}                                       & \multicolumn{2}{c}{QC-7B}                                       & \multicolumn{2}{c}{QC-14B}                                      \\ \midrule
Vul-type & {Acc} & {F1} & {Acc} & {F1} & {Acc} & {F1} \\ \midrule
22       & 0.6                                & 0.33                              & 0.8                                & 0.75                              & 0.8                                & 0.8                               \\
78       & 0.7                                & 0.57                              & 0.9                                & 0.89                              & 0.9                                & 0.91                              \\
79       & 0.6                                & 0.5                               & 0.9                                & 0.91                              & 1                                  & 1                                 \\
89       & 0.8                                & 0.75                              & 0.9                                & 0.89                              & 1                                  & 1                                 \\
125      & 0.6                                & 0.67                              & 0.6                                & 0.33                              & 0.6                                & 0.67                              \\
190      & 0.7                                & 0.57                              & 0.8                                & 0.83                              & 0.9                                & 0.89                              \\
416      & 0.8                                & 0.75                              & 0.7                                & 0.77                              & 0.8                                & 0.8                               \\
476      & 0.8                                & 0.83                              & 0.7                                & 0.67                              & 0.6                                & 0.71                              \\
787      & 0.6                                & 0.71                              & 0.8                                & 0.8                               & 0.8                                & 0.83                              \\ \midrule
Ave      & 0.69                       & 0.63                     & 0.79                       & 0.76                              & 0.82                       & 0.85        \\ \bottomrule          
\end{tabular}
\end{table}

\subsection{Boarder Impact}
Our work on code vulnerability detection and correction has significant societal implications. \textbf{Positively}, it enhances code security, potentially reducing data breaches and cyberattacks that impact millions of users annually. Secure code generation tools democratize cybersecurity expertise, benefiting resource-constrained organizations and critical infrastructure. \textbf{Negatively}, adversarial applications of our techniques could be used to deliberately introduce subtle vulnerabilities or automate exploitation of existing weaknesses. Additionally, over-reliance on automated security tools may create false confidence and reduce human oversight. We encourage responsible deployment with human-in-the-loop verification and recommend against using these methods in high-risk applications without thorough testing.

\end{document}